\documentclass[12pt]{article}

\usepackage[english]{babel}
\usepackage[intlimits]{amsmath}

\begin{document}
\title{Fractional Langevin equation to describe anomalous diffusion.\\}
\author{\begin{Large} V.Kobelev, E.Romanov \end{Large} \\\\
\emph{Institute of Metal Physics RAS, Ekaterinburg, 620219, RUSSIA}}
\date{25 May 1999}
\bigskip

\maketitle

\begin {center}   Abstract     \end{center}
A Langevin equation with a special type of additive random source is
considered. This random force presents a fractional order derivative of
white noise, and leads to a power-law time behavior of the mean square
displacement of a particle, with the power exponent being noninteger. More
general equation containing fractional time differential operators instead
of usual ones is also proposed to describe anomalous diffusion processes.
Such equation can be regarded as corresponding to systems with incomplete
Hamiltonian chaos, and depending on the type of the relationship between
the speed and coordinate of a particle yields either usual or fractional
long-time behavior of diffusion. Correlations with the fractional
Fokker-Planck equation are analyzed. Possible applications of the proposed
equation beside anomalous diffusion itself are discussed.
\newpage
\section{Introduction}

In recent years, growing attention has been focused on the processes that
take place in random disordered media, and in dynamic systems
demonstrating chaotic behavior. A special place here is taken by the
systems with incomplete Hamiltonian chaos, which trajectories in the phase
spacecan be portrayed as a set of "islands around islands" with
self-similar (fractal) structure. In such systems the lifetime of any
state of regular motion is also random. Incomplete chaos results in a few
interesting phenomena that, among others, include anomalous properties of
transport processes. That is associated with the fact that the islands of
stability in this case become to act as a system of  traps with a certain
given distribution of the trapping time. For example, well known is the
phenomenon of anomalous diffusion that is characterized by the time
function of the mean squared displacement of a diffusing particle, which
is described not by the Einstein's law but by an power function with
fractional exponent \cite{bou,kl}.
 $$ \langle
(\Delta x)^{2}\rangle \propto t^{\alpha}\;\;\;\; \alpha\neq1 $$ In most
cases, such behavior is considered to be connected with self-similar
properties of the diffusion medium. As this takes place, the Fokker-Planck
equation describing such diffusion process was shown to involve the
integrodifferential Riemann-Liouville operators $I^{\nu}$ and $D^{\nu}$ of
fractional order \cite{sa} (see Appendix A). Possible interpretation of
the physical meaning of these operators was offered in \cite{ni,za}, and
assumes that a system described by equations with fractional derivatives
or integrals possesses a "selective" memory that acts only in the points
within a set of dimensionality $\nu$, and is in accordance with the ideas
about a certain self-similar (say, Levy-type or fractal) distribution of
traps and waiting times \cite{za}. Fractional Fokker-Planck equations were
studied in \cite{za}-\cite{kol1}, and proved to be quite useful to model
anomalous diffusion processes. The most general form of such equations was
introduced in \cite{za,ko,kol1} and in a simple case looks
\begin{equation}
  \frac{\partial ^{\nu}}{\partial t^{\nu}}n=
  D\frac{\partial ^{2 \gamma}}{\partial x^{2 \gamma}}n \label{eq:frdif}
\end{equation}
where $n$ is the concentration of diffusing substance and the fractional
differential operators on the right-hand side and on the left-hand side
are either fractional Riemann-Liouville derivatives \cite{za}-\cite{ko} or
so called local fractal derivatives \cite{kol1}.

Another method of the anomalous transport properties description refers to
their microscopic motion and involves introducing fractional derivatives
into a stochastic process that characterizes the Brownian motion. As the
starting point to be changed the Wiener stochastic process that defines
the displacement of a particle with time is taken
$$ x-x_{0}=\int_{0}^{t}F(t^{\prime})dt^{\prime} $$
where $F(t)$ is the the Gaussian white noise (the brackets here mean
averaging over the possible realizations of the process),
 $$ \langle F(t) \rangle =0,\;\;\;\;\langle
F(t_{1})F(t_{2}) \rangle =q\delta (t_{1}-t_{2}) $$ and the fractal
Brownian motion is obtained by substituting the Riemann integral for a
fractional one \cite{man}-\cite{seb}
\begin{equation}
x-x_{0}=I^{\alpha}F(t)\equiv \frac{1}{\Gamma (\alpha)}
\int^{t}_{0}\frac{F(t^{\prime})}{(t-t^{\prime})^{1-\alpha }}dt^{\prime}
         \label{eq:frwin}    \end{equation}
The fractional integral of the white noise was named the
fractional noise. This noise is Gaussian but nonstationary. Its
correlation function has the form
\begin{eqnarray}
  &&\langle x(t_{1})x(t_{2})\rangle=\frac{t_{>}^{\alpha -1}t_{<}^{\alpha}}
  {\alpha \left[\Gamma(\alpha)\right]^{2}}\;
   _{2}F_{1}(1,1-\alpha;1+\alpha;t_{<}/t_{>})       \label{eq:frcorr} \\
    &&t_{<}=min(t_{1},t_{2}), \;\;   t_{>}=max(t_{1},t_{2}) \nonumber \\
    &&\langle x(t)x(t)\rangle \; \propto \; t^{2\alpha - 1} \nonumber
                        \end{eqnarray}
In principle, this noise can be made stationary, but only provided that
$-\infty$  is taken in Eq.(\ref{eq:frwin}) as the lower limit of
integration instead of 0. However, in this case one has to somewhat modify
the power kernel in order to provide the convergence of the integral. We
will briefly discuss this below, at the end of the Section 2.

Fractal Brownian motion is of great interest not only from mathematics and
theory of stochastic processes points of view but also in terms of
physical applications. It can be used in describing polymer chains,
electric transport in disordered semiconductors, diffusion on comb-like
structures, and so on.  However, it is well known that in many cases the
most convenient way of describing the Brownian diffusion of particles is
not the Wiener process but rather the Ornstein-Uhlenbeck process (or
Langevin method \cite{kl,ri,vkam}) that is based on the solution of
stochastic differential equation
\begin{equation}
  \frac{d}{dt}v=-\gamma v+F(t) \label{eq:lang} \end{equation}
where $v$ is the velocity of a Brownian particle, $\gamma$ means the
factor of liquid friction, and $F(t)$ is the random source characterizing
the properties of medium where diffusion occurs. Displacement of a
particle and its path are determined not directly but through the
integration of the instant velocities. Apart from the problem of Brownian
motion itself, this method is widely applied to describe various systems
subjected to external noise. This brings up the question of whether the
Langevin equation can be written for anomalous diffusion as well, and if
so, what will be the structure of the corresponding random source in it.
In principle, if the Fokker-Planck equation is known then the Langevin
equation can be derived from it. However, this method involves certain
difficulties, and gets even more complicated because of the intricate
structure of fractional integrodifferential operators.  So, we shall
postulate the form of such equation in further sections of our paper.
Actually, we shall consider three forms of such equation that have in
common the method of introducing the memory by means fractional
derivatives yet differ in the properties of medium and in the ways of
setting up the problem.

\section{  Langevin Equation with Fractional Derivatives}

\bf a. \rm  As it has been mentioned above, introducing the fractional
differential operators into the Fokker-Planck equation makes it possible
to describe the anomalous transport process quite correctly. Therefore,
let us consider the following equation that differs from the usual
Langevin equation by replacing the first derivative with respect to time
by the fractional derivative of order $\nu$
\begin{equation}
  \frac{d^{\nu}}{dt^{\nu}}v=-\gamma v+F(t) \label{eq:frl1d} \end{equation}
Applying the fractional integral operator to both left-hand and right-hand
sides of this equation, one has
 \begin{equation}
       v-v_{0}=_{0}I^{\nu}_{t}(-\gamma v+F(t)) \label{eq:frl1i} \end{equation}
Expressing the fractional integral in the explicit form, we rewrite it as
follows
\begin{equation}
   v=v_{0}+A(t)-\frac{\gamma}{\Gamma (\nu)}
    \int^{x}_{0}\frac{v(x^{\prime})}{(x-x^{\prime})^{1-\nu}}dx^{\prime},
    \;\;\; A(t)=\frac{1}{\Gamma (\nu)}
    \int^{x}_{0}\frac{F(x^{\prime})}{(x-x^{\prime})^{1-\nu}}dx^{\prime}
           \label{eq:frl1}             \end{equation}
This equation can be easily solved by standard techniques for Volterra
integral equations, and its solution has the form
\begin{equation}  \label{eq:sol0}
  v=v_{0}E_{1,\nu}(-\gamma t^{\nu})+
  \int^{t}_{0}F(t^{\prime})(t-t^{\prime})^{\nu-1}
   E_{\nu,\nu}[-\gamma(t-t^{\prime})^{\nu}]dt^{\prime}
                                         \end{equation}
where $E_{\alpha,\beta}(z)$ is so called Mittag-Leffleur function
\cite{dzh}
     $$ E_{\alpha,\beta}(z)=\sum^{\infty}_{k=0}\frac{z^{k}}{\Gamma
         (\alpha+\beta k)} $$
If $\nu=1$ then Eq.(\ref{eq:sol0}) reduces to the solution of the usual
Langevin equation \cite{ri}
 $$ v=v_{0}e^{-\gamma
t}+\int^{t}_{0}F(t^{\prime})e^{-\gamma(t-t^{\prime})}dt^{\prime} $$
If $F(t)$ is taken, as usual, to be a Gaussian $\delta$-correlated source
with zero mean, then the velocity correlation function has the form
                          \begin{eqnarray}
\nonumber \langle v(t_{1})v(t_{2}) \rangle &=&v^{2}_{0}E_{1,\nu}(-\gamma
t^{\nu}_{1}) E_{1,\nu}(-\gamma t^{\nu}_{2})  \\
&+&\!q\int_{0}^{min(t_{1},t_{2})}dt^{\prime}\frac{E_{\nu,\nu}\bigl[-\gamma
(t_{1}-t^{\prime})^{\nu}\bigr]}{(t_{1}-t^{\prime})^{1-\nu}}
\frac{E_{\nu,\nu}\bigl[-\gamma (t_{2}-t^{\prime})^{\nu}\bigr]}
{(t_{2}-t^{\prime})^{1-\nu}}           \label{eq:vkorr1}
                           \end{eqnarray}
The integral from the right-hand side can be taken in the explicit form
only if we introduce $t_{>}$ and $t_{<}$ from Eq.(\ref{eq:frcorr}), which
is rather inconvenient. However, we can obtain several important results
without having to calculate the integral directly. It is only essential
that the integral is a symmetrical function of its arguments $t_{1}$ and
$t_{2}$. Further, let us find the corresponding mean squared displacement
of the particle, which motion is described by the equation
Eq.(\ref{eq:frl1d}). It follows the expression
\begin{equation}
   \langle (\Delta x)^{2}\rangle
  =\int_{0}^{t}\int_{0}^{t}\langle v(t_{1})v(t_{2}) \rangle dt_{1}dt_{2}
              \label{eq:dx}                        \end{equation}
which, as shown in the Appendix B, for correlation function of
Eq.(\ref{eq:vkorr1}) can be reduced to
\begin{eqnarray}
\nonumber \langle (\Delta x)^{2}\rangle&=&v^{2}_{0} \Bigl[
tE_{2,\nu}(-\gamma t^{\nu}) \Bigr]^{2} \\ &+&\frac{q}{\gamma ^{2}}
\biggl(t-2tE_{2,\nu}(-\gamma t^{\nu})+\int_{0}^{t}\Bigl[E_{1,\nu}(-\gamma
t^{\nu}) \Bigr]^{2}dt \biggr)
              \label{eq:dx1_1}                       \end{eqnarray}
At large values of the argument $E_{1,\nu}(-\gamma t^{\nu}) \propto
E_{2,\nu}(-\gamma t^{\nu}) \propto 1/\gamma t^{\nu}$. Therefore, the
integral in Eq.(\ref{eq:dx1_1}) necessarily converges at $t \to
\infty$ for $1/2<\nu$, and the leading term in the asymptotics  is
 \begin{equation}
     \langle (\Delta x)^{2}\rangle=\frac{q}{\gamma ^{2}}t
      \label{eq:dx1_2}          \end{equation}
just as it is the case for the classical Langevin equation with the
first derivative with respect to time. Even when $\nu<1/2$ and the
integral diverges, it increases more slowly then the first term and
the linear asymptotics remains. Leaving the discussion of this point
for later, we shall try to modify somehow the initial equation.

\bf b. \rm First, it must be noted that in Eq.(\ref{eq:dx1_1}) we assumed
that, as usually, the velocity $v$ was defined as the first derivative of
the coordinate with respect to time and, therefore,
$x=\int_{0}^{t}v(t)dt$. Actually, it makes sense to consider a more
general relationship
\begin{equation}
      x=\frac{1}{\Gamma (\nu)}\int_{0}^{t}\frac{v(t^{\prime})}
     {(t-t^{\prime})^{1-\nu}}dt^{\prime}
             \label{eq:x_nu}            \end{equation}
that corresponds to the most complete possible description of the system
memory by means of the fractional integrals - the particle's displacement
is defined by velocity only in the points within a time interval of
dimension $\nu$. Though in this case the physical meaning of the
corresponding inverse definition of velocity as a fractional derivative of
the coordinate with respect to time needs to be explained. In order to
clear this point, let us recollect that microscopic motion of a diffusing
particle represents a twisted and everywhere nondifferentiable curve. For
such curve, however, one can often find a derivative of fractional order
\cite{kol2} that, in fact, is a  derivative of the trajectory, averaged
with a power weight, and the observed motion of the particle is the thus
averaged motion. In terms of memory, it means that for fractal paths
(which are even the paths of classical Brownian particle) some of the instant
velocities and displacements do not contribute into the resulting
macroscopic motion. As this occurs, the behavior of the solution changes,
and diffusion becomes anomalous. In this case instead of Eq.(\ref{eq:dx1_1})
we obtain
\begin{eqnarray}
\langle (\Delta x)^{2}\rangle &=&
_{0}I_{t_{1}}^{\nu}\:_{0}I_{t_{2}}^{\nu}\langle v(t_{1})v(t_{2}) \rangle=
 v^{2}_{0} \frac{\Bigl[E_{1,\nu}(-\gamma
t^{\nu})-1\Bigr]^{2}}{\gamma ^{2}} \nonumber \\ &+& \frac{2qt^{2\nu
-1}}{\gamma^{2}}\sum_{k,l=1}^{\infty}\frac{(-\gamma)^{k+l}}{\Gamma (\nu
+\nu k)\Gamma (\nu l)} \frac{\Gamma (\nu k+\nu l+\nu -1)}{\Gamma (\nu
k+\nu l+2\nu)}t^{\nu k+\nu l}      \label{eq:dx2_2}
\end{eqnarray}
At large times the first term tends to $v^{2}_{0}/ \gamma^{2}$, and the
series in the second term converges (see Appendix C). Therefore we finally
have
\begin{equation}
    \langle (\Delta x)^{2}\rangle \propto N\frac{q}{\gamma ^{2}}
     t^{2\nu -1}, \;\;\;\;  1<N<\frac{2}{\Gamma (\nu)}
            \label{eq:dx2_3}                      \end{equation}
that coincides with the asymptotics obtained in \cite{ko,kol1} for
the generalized equation of fractal diffusion Eq.(\ref{eq:frdif}) if
$\gamma=1$, and with the asymptotics of fractional Brownian motion
\cite{man}-\cite{seb}. The larger $\nu$ the faster the particle
moves. More accurate estimations of factor $N$ can be obtained either
following the method similar to that stated in the Appendices B and C
(which in this case would be quite a challenge) or numerically.

\bf c. \rm Let us now investigate separately the influence of memory and
fractal behavior on the regular and random components of the force acting
on a particle. It turns out that the same fractal asymptotic behavior can
be demonstrated by taking into account only the memory for the random
force component in the initial Langevin equation, which has a more clear
physical meaning (cf. Eq.(\ref{eq:frl1})):
  \begin{equation}
       v=v_{0}-\gamma \int_{0}^{t}v(t^{\prime})dt^{\prime}+
    \frac{1}{\Gamma (\nu)}\int_{0}^{t}\frac{F(t^{\prime})}{(t-t^{\prime})^{\nu}}dt^{\prime}
         \label{eq:frl3}                 \end{equation}
In fact, it means that in Eq.(\ref{eq:lang}) the random source is not
$\delta$-correlated but represents the fractional derivative of the
$\delta$-correlated process $g(t)$, which can be taken in the sense of the
generalized functions  \cite{ge}
 \begin{equation}
    F(t)=\: _{0}D_{t}^{1-\nu}g(t) \;\;\;\;\;\;\langle g(t_{1})g(t_{2})
     \rangle =q\delta(t_{1}-t_{2})
            \label{eq:frnoise}                  \end{equation}
Similar equation was actually considered in \cite{shao}. It was shown
numerically there that when in the Langevin equation Gaussian white
noise is replaced by the fractional Gaussian noise \cite{man,fe} it
can yield the spectra for homogeneous Eulerian and Lagrangian
turbulence. However, equation Eq.(\ref{eq:frl3}) can be easily solved
analytically, and the solution reads
\begin{equation}
   v=v_{0}e^{-\gamma t}+
  \int^{t}_{0}F(t^{\prime})(t-t^{\prime})^{\nu-1}
   E_{\nu,1}\bigl[-\gamma(t-t^{\prime})\bigr]dt^{\prime}
      \label{eq:frl3sol}                            \end{equation}
The corresponding expression for correlation is
 \begin{eqnarray}
\nonumber \langle v(t_{1})v(t_{2}) \rangle &=&v^{2}_{0}e^{-\gamma
(t_{1}+t_{2})}  \\ &+&
q\int_{0}^{min(t_{1},t_{2})}dt^{\prime}\frac{E_{\nu,1}\bigl[-\gamma
(t_{1}-t^{\prime})\bigr]}{(t_{1}-t^{\prime})^{1-\nu}}\frac{E_{\nu,1}\bigl[-\gamma
(t_{2}-t^{\prime})\bigr]}{(t_{2}-t^{\prime})^{1-\nu}}
            \label{eq:vkorr2}                        \end{eqnarray}
Calculation of the mean squared displacement is similar to that in the
Appendix B and results in
\begin{eqnarray}
  & &\langle (\Delta x)^{2}\rangle= v^{2}_{0}\frac{\bigl(1-e^{-\gamma t}\bigr)^{2}}
   {\gamma ^{2}}+\frac{q}{\gamma ^{2}}S \nonumber \\
  & &S=2\sum_{k,l=0}^{\infty}\frac{(-\gamma )^{k+l}}{\Gamma (1+\nu +k)\Gamma(\nu+l)}
  \frac{t^{k+l+2\nu +1}}{(k+l+2\nu)(k+l+2\nu +1)}
       \label{eq:dx3_1}            \end{eqnarray}
By differentiating the sum two times, we obtain $$
\frac{d^{2}}{dt^{2}}S=\frac{d}{dt}
     \Bigl[t^{\nu}E_{1+\nu,\nu}(-\gamma t)-1 \Bigr]^{2} $$
from which it follows that when $t\to\infty$ it is the second term that
defines the asymptotic behavior of the displacement
\begin{equation}
\langle (\Delta x)^{2}\rangle \propto \frac{1}{(2\nu
-1)\bigl[\Gamma(\nu+1)\bigr]^{2}}\frac{q}{\gamma ^{2}} t^{2\nu -1}
             \label{eq:dx3_2}                        \end{equation}
which agrees with Eq.(\ref{eq:dx2_3}) up to the constant multiplier.

\bf d. \rm Finally, let us consider the third case, that is when the
memory is taken into account only for the friction force
\begin{equation}
 v=v_{0}-\gamma \frac{1}{\Gamma(\nu)}\int_{0}^{t}\frac{v(t^{\prime})}
  {(t-t^{\prime})^{\nu}}dt^{\prime}+ \int_{0}^{t}v(t^{\prime})dt^{\prime}
     \label{eq:frl4}                             \end{equation}
where $F(t)$ is again the Gaussian white noise (cf. Eqs.(\ref{eq:frl1}),
(\ref{eq:frl3})). It means that now it is the dissipative force that is
proportional to the fractional derivative of velocity
$f_{diss}=\,_{0}D_{t}^{1-\nu}v(t)$. Its solution is
\begin{equation}
   v=v_{0}E_{1,\nu}(-\gamma t^{\nu})+
\int^{t}_{0}F(t^{\prime})E_{\nu,\nu}\bigl[-\gamma(t-t^{\prime})^{\nu}\bigr]
  dt^{\prime}
            \label{eq:frl4sol}                  \end{equation}
The contribution from the second term in the mean squared displacement
results in the power function of time in the following form
\begin{equation}
    \langle (\Delta x)^{2}\rangle \propto M\frac{q}{\gamma ^{2}}
              t^{3-2\nu}, \;\;\;\; 1<M<\frac{1}{\Gamma (1+\nu)}
              \label{eq:dx4_1}                   \end{equation}
It should be noted here that, unlike the previous two cases, here no
limitations are imposed on the value of $\nu$, and depending on
whether it is greater or less than 1, we can obtain either
subdiffusion with the power exponent less than 1, or superdiffusive
behavior with the power exponent greater than 1, observed, for
instance, for a phase for wave propagation in nonlinear or random
media. The greater is $\nu$, the slower the particle moves (it was
the other way around in the previous cases), because at $\nu>1$ the
corresponding dissipative force is no longer a fractional derivative
of the velocity but a fractional integral, and, hence, its influence
increases.  At $\nu>3/2$, the mean squared displacement
Eq.(\ref{eq:dx4_1}) tends to zero with time increasing, which
corresponds to a stop in the particle motion since the terms
neglected in Eq.(\ref{eq:dx4_1}), as well as in
Eqs.(\ref{eq:dx2_2})-(\ref{eq:dx2_3}), are either vanishingly small
or constant, and means that particle energy dissipates faster than it
is pumped. On the other hand, limitations $1/2<\nu<1$ were used in
the previous cases only to simplify the estimations for the sums
which were shown to be convergent and therefore in principle it must
not affect the main results and they also can give sub- and
superdiffusion behavior. But because of the fact that prehistory
influences only on the dissipative force, now diffusion is impeded
when we increase the value of $\nu$.

Let us now analyze the results obtained.  If we substitute the first
derivative in the Langevin equation for the fractional one but use the
same relationship for velocity and coordinate, then the solution possesses
the same linear asymptotic behavior as the initial solution does.  This
means that taking into account the memory for the friction force and
random force at the same moments of time does not affect the particle's
motion at large times but only in the beginning of the motion.  In a
sense, this is quite reasonable, since a random source provides a particle
with an additional energy, and friction results in its dissipation.  At
larger times, when the system arrives at the stable state, these two
processes compensate each other, but only provided that they have the same
duration.

Using the relationship Eq.(\ref{eq:x_nu}) means that the particle's motion
equation written in the Newtonian form reads
\begin{equation}
  \frac{d^{2\nu}x}{dt^{2\nu}}=\frac{d^{\nu}x}{dt^{\nu}}+F(t)
                         \end{equation}
and is the most general in the sense of memory calculation - every
derivative with respect to time becomes fractional. Then anomalous
properties of the diffusion arise from fractional integration of velocity,
that is, from the assumption only part of instant velocities contribute to the
final path.

The anomalous diffusion is also the result of the Langevin equation with a
source that is a fractional derivative of the white noise.  The
autocorrelation function of such noise, just like that in
Eq.(\ref{eq:frwin}), has a power behavior yet negative values of the
exponent and therefore appears to be a generalization of the flicker-noise
(see \cite{kl}). It should be noted that a stationary noise can be
obtained only by using the Liouville fractional derivative with the
infinite lower limit. However, when used directly in the equations
Eqs.(\ref{eq:frl1}), (\ref{eq:frl3}), (\ref{eq:frl4}), these derivatives
would lead to the divergences because of the power integrands. Besides,
this is not quite reasonable, since in real situations there is always a
moment at which the motion starts, and assuming this moment as infinitely
distant is possible only in a system where characteristic relaxation
scales exist. Fractional Brownian motion, however, can not be assigned to
such processes - its memory is described by a power function and leads to
the long-range correlations having no time scale of their own.
Nevertheless, the nonstationary of the noise should not lead to a
misunderstanding. As we shall illustrate below, such process can be
regarded as stationary only in a wide sense (the corresponding transition
probabilities depend uniquely upon the duration of the transition $\Delta
t$) only for small $\Delta t$, that is for $\Delta t<<t$.

The absence of stationarity is an important and rather obvious property of
anomalous diffusion, although it is not always given proper consideration.
Here it is relevant to note the following. Normally, the Brownian motion
can be described with two different stochastic processes. The
Ornstein-Uhlenbeck process is strictly stationary but does not have
independent increments. Moreover, its increments are not even
uncorrelated. The Wiener process, which is the integrated
Ornstein-Uhlenbeck process in the limit of intense friction and noise, has
stationary independent increments, but is neither strictly stationary nor
a wide-sense stationary. The process under investigation with a fractional
derivative of the white noise seems to be an intermediate process that
does not have stationary increments but is asymptotically stationary in a
wide sense. The fractal nature of motion leads to the fact that even this
kind of stationarity is observed only within short periods of time, which
means that the process becomes quasistationary. This conclusion is in a
good agreement with the notion that anomalous diffusion is just an
intermediate asymptotic behavior for systems of certain types.

Finally, if the prehistory affects only the dissipative force acting on a
particle, its behavior also becomes anomalous.

\section{Probability distributions}

Let us now study the probability distributions that arise from the
stochastic processes described above.  For the sake of convenience, let us
consider only the process with the noise that is a fractional derivative
of the white noise Eq.(\ref{eq:frl3}). The other cases are considered in
perfect analogy to this one and give similar results.

The distribution of the particle coordinate will be Gaussian. It follows
from the fact that the equations are linear both with respect to the
external additive noise and the stochastic variable. In order to determine
the transition probability  $P(x,t+\Delta t;y,t)$ let us consider
transition moments at large times
\begin{equation}
  M_{n}=\vert x-y \vert^{n}=\int_{t}^{t+\Delta t}...\int_{t}^{t+\Delta t}
  \langle v(t_{1})...v(t_{n})\rangle dt_{1}...dt_{n}=
  \begin{cases} 0,& n=2k+1 \\ \frac{(2k)!}{2^{k}k!}I^{k},&n=2k \end{cases}
           \label{eq:mom}                \end{equation}
where
\begin{eqnarray}
 I&=&\int_{t}^{t+\Delta t}\int_{t}^{t+\Delta t}
     \langle v(t_{1})v(t_{2})\rangle dt_{1}dt_{2} \nonumber \\
 &=&\int_{t}^{t+\Delta t}dt_{1}\int_{t}^{t+\Delta t} dt_{2}
     \int_{0}^{min(t_{1},t_{2})}
     \frac{E_{\nu,1}\bigl[-\gamma (t_{1}-t^{\prime})\bigr]}
           {(t_{1}-t^{\prime})^{1-\nu}}
     \frac{E_{\nu,1}\bigl[-\gamma (t_{2}-t^{\prime})\bigr]}
          {(t_{2}-t^{\prime})^{1-\nu}}dt^{\prime}
       \label{eq:disp}             \end{eqnarray}
This implies that the probability of transition obeys the Gaussian
distribution with dispersion $I$.  In order to identify time dependence of
the dispersion, let us rewrite the internal integral in the form
$\int_{0}^{t+\Delta t}=\int_{0}^{t}+ \int_{t}^{t+\Delta t}$. In this case
Eq.(\ref{eq:disp}) will have two terms, the first of which allows changing
the order of integration since the limits of integration are no longer
dependent on the intrinsic variables:
$$ I_{2}=\int_{0}^{t}dt^{\prime}\Biggl[\int_{0}^{t+\Delta t}
   \frac{E_{\nu,1}\bigl[-\gamma(t_{1}-t^{\prime})\bigr]}
        {(t_{1}-t^{\prime})^{1-\nu}}dt_{1}\Biggr]^{2}  $$
At $\Delta t<<t$ we then get
\begin{equation}
 I_{2}=A\frac{q}{\gamma ^{2}}\frac{\Delta t^{2}}{t^{3-2\nu}}, \;\;\;\;\;\;
   \frac{1}{2\nu \bigl[\Gamma(\nu)\bigr]^{2}}<A<
    \frac{1}{(2\nu-1) \bigl[\Gamma(\nu)\bigr]^{2}}
     \label{eq:sec_term}               \end{equation}
The second terms that appeares in Eq.(\ref{eq:disp}) after the substitution
$t_{1}\to t_{1}-t,\, t_{2}\to t_{2}-t,\, t^{\prime} \to t^{\prime}-t$
reduces to the mean squared displacement given by Eq.(\ref{eq:dx3_2}) during
time $\Delta t$
$$ I_{1}=\frac{1}{(2\nu -1)\bigl[\Gamma(\nu+1)\bigr]^{2}}\frac{q}
   {\gamma ^{2}} t^{2\nu -1}  $$
which is much greater than $I_{2}$ if $\Delta t<<t$.  Hence,
\begin{eqnarray}
  P(x,t+\Delta t;y,t)&=&\sqrt{\frac{\gamma^{2}(2\nu-1)\bigl[\Gamma(\nu+1)\bigr]^{2}}
  {2\pi q }\frac{1}{\Delta t^{2\nu-1}}} \nonumber \\
  & &\times \exp \Biggl(-\frac{\gamma^{2}(2\nu-1)\bigl[\Gamma(\nu+1)\bigr]^{2}}{2q }
     \frac{\vert x-y \vert ^{2}}{\Delta t^{2\nu-1}}\Biggr)
      \label{eq:probxy}            \end{eqnarray}
This expression also gives a "suitable" result for the coordinate
distribution $W(x,t)$ of the wandering particle that agrees with
\cite{kl,fe,man,seb}
\begin{eqnarray}
  W(x,t)&\equiv& P(x,t;0,0)=\frac{1}{\sqrt{2\pi B t^{2\nu-1}}}
   \exp \Biggl(-\frac{x^{2}}{2B t^{2\nu-1}}\Biggr)
           \label{eq:probx0}  \\
  B&=&\frac{1}{(2\nu -1)\bigl[\Gamma(\nu+1)\bigr]^{2}}\frac{q}
  {\gamma ^{2}} \nonumber
              \end{eqnarray}
and follows from the fact that if we take $y=0, t_{0}=0$ as the initial
point of transition, then the $I_{2}=0$ holds disregarding the value of
$\Delta t$. Therefore, at large times the described process becomes
quasistationary in a wide sense, that is, the probability of transition
during a small time interval $\Delta t$ depends on the value of $\Delta t$
only. However, if the condition $\Delta t<<t$ is not fulfilled, then
$I_{2}$ can no longer be given with the simple expression like
Eq.(\ref{eq:sec_term}) and thus can not be neglected as compared to $I_{1}$.
It should also be noted that such process on any time scales is not
Markovian  since the equations Eqs.(\ref{eq:probxy})-(\ref{eq:probx0}) do
not comply with the Chapmen-Kolmogorov equation
$$  P(x,t+\Delta t)=\int P(x,t+\Delta t;y,t)P(y,t)dy  $$
which can be verified by direct substitution. Indeed, if we introduce in
Eq.(\ref{eq:probx0}) new time variable as $\tau=t^{2\nu-1}$, then the
Markovian Gaussian process will be that with the transition probability
distribution
$$ P(x,\tau+\Delta \tau;y,\tau)=\frac{1}{\sqrt{2\pi D \Delta\tau^{2\nu-1}}}
  \exp \Biggl(-\frac{\vert x-y \vert ^{2}}{2D \Delta\tau^{2\nu-1}}\Biggr)
  \Delta\tau=(t+\Delta t)^{2\nu-1}-t^{2\nu-1} $$
and at small $\Delta t $
$$ \Delta \tau \approx (2\nu-1)t^{2\nu-2}\Delta t $$
It is worth to note that in this case the property of wide- sense
stationarity is lost for all time scale.

Non-Markovity is a consequence of the memory of the past process
existence. If for a Markovian process the future is uniquely determined by
the present, then in the initial equation Eq.(\ref{eq:frl3}) the behavior of
a particle in the next moment of time is, generally, dependent on the
whole previous history starting from the very beginning of the motion, and
the transition probability $P(x,t+\Delta t; y, t)$ does not depend on the
time $t$ only when $\Delta t \to 0$. In this connection the question of
transition from Eq.(\ref{eq:frl3}) to the Fokker-Planck-type equation still
remains to be solved. Indeed, the proposed in \cite{za,kol1} method is
based on the fractional Taylor series expansion of the transition
probabilities in the Chapmen-Kolmogorov equation. For example, according
to \cite{za}, the fractional Fokker-Planck equation  for
the transition probability given by Eq.(\ref{eq:probxy}) takes the form
\begin{equation}
  \frac{\partial ^{2\nu-1}}{\partial t^{2\nu-1}}W(x,t)=D\frac {\partial ^{2}}
   {\partial x^{2}}W(x,t)
        \label{eq:frdif1}            \end{equation}
Solution of this equation was obtained in \cite{ko,po} and shown to be
\begin{equation}
  W_{1}(x,t)=\frac{1}{\sqrt{Dt^{2\nu-1}}}H_{1,1}^{1,0}
   \left( \frac{x^{2}}{Dt^{2\nu-1}} \vert \begin{array}{cc}(1,\nu-1/2)\\
       (1,1) \end{array} \right)
             \label{eq:probx0_1}               \end{equation}
where $H_{1,1}^{1,0}$ is the so-called Fox function \cite{pr,mat}.
However, according to Eq.(\ref{eq:probx0})
\begin{equation}
   W(x,t)=\frac{1}{\sqrt{2\pi D t^{2\nu-1}}}
   \exp\left(-\frac{x^{2}}{Dt^{2\nu-1}} \right)
   = H_{0,1}^{1,0}\left( \frac{x^{2}}{Dt^{2\nu-1}} \vert \begin{array}{cc} - \\
     (1,1) \end{array} \right) \neq W_{1}(x,t)
          \label{eq:probx0_2}               \end{equation}
Therefore, the solution Eq.(\ref{eq:probx0_2}) does not satisfy
Eq.(\ref{eq:frdif1}) though the time and the coordinate of a particle
appears in its solution in the "correct" combination $x^{2}/t^{2\nu
-1}$. However, as the stochastic process under consideration is not
globally stationary, this fact seems to influence the structure of
the corresponding Fokker-Planck equation. In particular, such
equation may also include the terms that characterize the sources and
drains of probability, and therefore its structure will no longer be
as simple as in Eq.(\ref{eq:frdif1}). Probably this problem may be
solved by studying the waiting time probabilities and then writing
down the equations similar to those used in \cite{chuk}.

The average kinetic energy of a particle can be obtained as
\begin{equation} \label{eq:Ekin1}
  \langle E \rangle = \langle [v(t)]^{2} \rangle
\end{equation}
Substituting here the expression for the velocity correlation
Eq.(\ref{eq:vkorr2}) and taking into account that at large times, when the
motion becomes stable, the particle "forgets" the initial velocity and we
can neglect the first term in it (see also Eq.(\ref{eq:mom})), yields
\begin{equation}\label{eq:Ekin2}
 \langle E \rangle
 =\frac{mq}{2}\int_{0}^{t}\tau^{2\nu-2}\left[E_{\nu,1}(-\gamma
 \tau)\right]^{2}dx
\end{equation}
Since when $t \to \infty \;\; E_{\nu,1}(-\gamma \tau) \propto 1/\gamma
\tau$ and in the vicinity of zero the Mittag-Leffleur function is bounded,
for $1/2<\nu$ the integral in Eq.(\ref{eq:Ekin2}) converges. Therefore
(see \cite{ri}, the velocity of the moving particle will obey the
Maxwellian distribution.

Probability distribution for the processes described in the
paragraphs \bf 2b \rm (Eqs.(\ref{eq:frl1}) and (\ref{eq:x_nu}))  and
\bf 2d \rm (Eq. (\ref{eq:frl4})) are obtained in the same way.
Velocity distribution obeys the Maxwellian law, and the coordinate
distribution $W(x,t$) and transition probability $P(x,t+\Delta
t;y,t)$ at large times $t$ and small $\Delta t$ follow the Gaussian
law with the dispersion determined by the mean squared displacement
over times $t$ and $\Delta t$, respectively.

\section{Conclusion}

To conclude, we would like to note the following.  The equations and
models proposed in this paper are quite general and, represent a way to
introduce and describe a certain class of Gaussian non-Markovian
stochastic processes. Processes of this type are featured with the
"selective" memory acting only in the moments of time distributed over a
Cantor-type fractal set and taken into account by means of fractional
derivatives, and show promise in describing stochastic processes in
fractal media and systems with incomplete Hamiltonian chaos. One of such
processes is anomalous diffusion, observed in a wide variety of systems.
The method developed in this paper is an alternative to that utilizing the
fractional Fokker-Planck equations, although we could not trace fully the
relation between these two approaches and it still remains to be explored.
Nevertheless, the two approaches give similar results, and choosing
between them should probably be determined by a type of problem to be
solved and by practical convenience of calculations. Earlier the fractal
Brownian motion was represented only as a Wiener stochastic process of
fractional order (the integral of  order $\nu$ from the white noise or,
which is the same, the integral of the first order of the noise which is a
fractional derivative of order $\nu-1$ of the white noise). As shown in
the present paper, the Ornstein-Uhlenbeck process with fractional noise
leads to similar results and allows to obtain more general probability
distributions in an easier way, as compared to the path integrals
\cite{seb}. The equations proposed can also be easily generalized by
adding linear and nonlinear terms, and be used for calculation of various
statistic characteristics of real stochastic dynamic systems.

\appendix
\numberwithin{equation}{section}
\newpage
\section{Fractional dervatives}

There are about two dozens of different definitions for fractional
derivatives that are in one way or another adapted to various
features of classes of functions for which they are defined.  The
most comprehensive description of the mathematical aspects of the
issue is given in the monograph \cite{sa}. In physics the
Riemann-Liouville fractional derivative is the most commonly used.
Its definition goes back to the well-known Cauchy formula for
multiple integrals $$ _{a}I_{t}^{n}f(t)=\underset{n \mbox {
\footnotesize{times}}}{\underbrace{\int_{a}^{t}dt...\int_{a}^{t}dt}}
f(t)
 =\frac{1}{(n-1)!}\int_{a}^{t}dt^{\prime}\frac{f(t^{\prime})}{(t-t^{\prime})^{1-n}}
$$
Substituting the factorial for the Euler gamma-function, we can
generalize the formula by introducing the fractional exponent $\alpha$ as
follows
\begin{equation}
_{a}I_{t}^{\alpha}f(t)\equiv \frac{1}{\Gamma(\alpha)}
  \int_{a}^{t}dt^{\prime}\frac{f(t^{\prime})}{(t-t^{\prime})^{1-\alpha}}
\label{eq:frint}     \end{equation}
This expression is referred to as the
Riemann-Liouville fractional integral.  The fractional derivative is then
defined as an ordinary derivative of the integral of fractional order
\begin{eqnarray}
   _{a}D_{t}^{\alpha}f(t)&\equiv& \frac{d ^{\alpha}}{d t^{\alpha}}f(t)=
   \frac{d^{n}}{dt^{n}}\,_{a}I_{t}^{n-\alpha}f(t) \nonumber \\
   && =\frac{1}{\Gamma (n-\alpha)}
   \frac{d^{n}}{dt^{n}} \int^{t}_{a}
   \frac{f(t^{\prime})}{(t-t^{\prime})^{1+\alpha-n}}dt^{\prime} \;\;\;\;
     n-1 \le \alpha <n
\label{eq:frderiv}     \end{eqnarray} Fractional derivative can also be
treated in the form of convolution with power function, and in this sense
the definitions in Eqs.(\ref{eq:frint})-(\ref{eq:frderiv}) can be easily
transferred onto generalized functions.

\section{Calculation of the mean squared displacement}

The mean squared displacement of a particle during time $t$ is defined as
    \begin{equation}
\langle (\Delta x)^{2}\rangle =\int_{0}^{t}\int_{0}^{t}\langle
v(t_{1})v(t_{2})
 \rangle dt_{1}dt_{2} =v^{2}_{0} \Bigl[
 \int_{0}^{t}E_{1,\nu}(-\gamma t^{\prime~\nu})dt^{\prime} \Bigr]^{2}
 +\frac{q}{\gamma^{2}}I        \end{equation}
where the integral in square brackets in the first term equals to
$tE_{2,\nu}(-\gamma t^{\nu})$, and the second terms takes the form (we
expand the Mittag-Lieffleur function in a series)
\begin{eqnarray}
\nonumber I&=&\sum^{\infty}_{k,l=1}\frac{(-\gamma)^{k+l}}{\Gamma(\nu
k)\Gamma (\nu l)}
\int_{0}^{t}dt_{1}\int_{0}^{t}dt_{2}\int_{0}^{min(t_{1},t_{2})}
(t_{1}-t^{\prime})^{\nu k-1}(t_{2}-t^{\prime})^{\nu l-1}dt^{\prime} \\
&=&\sum^{\infty}_{k,l=1}\frac{(-\gamma)^{k+l}}{\Gamma(\nu k)\Gamma (\nu
l)} I_{kl}
                         \end{eqnarray}
Since the triple integral here does not change its form when interchanging
$t_{1} \leftrightarrow t_{2}$, then, assuming for certainty $t_{1}<t_{2}$,
we can write it as
 $$
I_{kl}=2\int_{0}^{t}dt_{2}\int_{0}^{t_{2}}dt_{1}\int_{0}^{t_{1}}dt^{\prime}
(t_{1}-t^{\prime})^{\nu k-1}(t_{2}-t^{\prime})^{\nu l-1}                $$
This integral is now easy to calculate using the rules of fractional
integration \cite{sa} and the properties of Gauss generalized
hypergeometric functions $_{2}F_{1}$  \cite{pr}
 \begin{eqnarray}
  \nonumber
I_{kl}&=&2\int_{0}^{t}dt_{2}\int_{0}^{t_{2}}dt_{1}\Gamma (\nu
k)_{0}I_{t_{1}}^{\nu k}(t_{2}-t_{1})^{\nu l-1} \\
  \nonumber
&=&\frac{2\Gamma (\nu k)}{\Gamma (\nu k+1)} \int_{0}^{t}dt_{2}t_{2}^{\nu
k+\nu l}\int_{0}^{t_{2}} \Bigl(\frac{t_{1}}{t_{2}}\Bigr)^{\nu
k}\:_{2}F_{1}\bigl( 1,1-\nu l;1+\nu k; \frac{t_{1}}{t_{2}}\bigr) d
\frac{t_{1}}{t_{2}} \\ &=&\frac{2}{\nu k(\nu k+\nu l)}
\int_{0}^{t}dt_{2}t_{2}^{\nu k+\nu l}=\frac{2 t_{2}^{\nu k+\nu l+1}}{\nu
k(\nu k+\nu l)(\nu k+\nu l+1})
                                      \end{eqnarray}
Substituting in  Eq.(\ref{eq:dx1_1}) yields
\begin{equation}
   I_{kl}=2t\sum^{\infty}_{k,l=1}\frac{(-\gamma t^{\nu})^{k+l}}
     {\Gamma (1+\nu k) \Gamma (\nu l)(\nu k+\nu l)(\nu k+\nu l+1)}
                                       \end{equation}
In order to evaluate the asymptotic behavior of this series at $t \to
\infty$, let us note that
 \begin{eqnarray}
    \nonumber
\frac{d^{2}I}{dt^{2}}&=& 2 \sum^{\infty}_{k,l=1}\frac{(-\gamma
t^{\nu})^{k}}{\Gamma (1+\nu k)} \sum^{\infty}_{k,l=1}\frac{(-\gamma
t^{\nu})^{l-1}}{\Gamma (\nu l)} \\
&=&\frac{d}{dt}\Bigl[\sum^{\infty}_{k=1}\frac{(-\gamma t)^{\nu k}}{\Gamma
(1+\nu k)} \Bigr]^{2}=\frac{d}{dt}\Bigl[ (E_{1,\nu}(-\gamma
t^{\nu})-1\Bigr]^{2}
                                       \end{eqnarray}
Integrated twice, and taken into account that
$I(t=0)=\frac{dI}{dt}|_{t=0}$, for the mean squared displacement it gives
\begin{eqnarray}
\nonumber \langle (\Delta x)^{2}\rangle&=&v^{2}_{0} \Bigl[
tE_{2,\nu}(-\gamma t^{\nu}) \Bigr]^{2} \\ &+&\frac{q}{\gamma ^{2}}
\biggl(t-2tE_{2,\nu}(-\gamma t^{\nu})+\int_{0}^{t}\Bigl[E_{1,\nu}(-\gamma
t^{\nu}) \Bigr]^{2}dt \biggr)
                                         \end{eqnarray}

\section{The sum evaluation}

We can find the asymptotic behavior of the double sum in
Eq.(\ref{eq:dx2_2}) (we designate it as $S_{1}$) in the following way.
First, let us find the upper boundary for this sum. To do this we
substitute the second cofactor $(\nu>1/2,~k,l\ge 1)$
        \begin{eqnarray}
  \nonumber
\frac{\Gamma (\nu k+\nu l+\nu -1)}{\Gamma (\nu k+\nu l+2\nu)} &=&
\frac{\Gamma (\nu k+\nu l+\nu)}{(\nu k+\nu l+\nu -1)\Gamma
 (\nu k+\nu l+2\nu)} \\
  &<& \frac{1}{\nu k+\nu l+\nu -1} <\frac{1}{\nu l}
                                                      \end{eqnarray}
hence
                  \begin{equation}
S_{1}<\sum_{k=1}^{\infty}\frac{(-\gamma t^{\nu})^{k}}{\Gamma (\nu +\nu k)}
      \sum_{l=1}^{\infty}\frac{(-\gamma t^{\nu})^{l}}{\Gamma (1+\nu l)}
   = \bigl(E_{\nu ,\nu}(-\gamma t^{\nu})-\frac{1}{\Gamma (\nu)}\bigr) \bigl
   (E_{1 ,\nu}(-\gamma t^{\nu})-1\bigr)
                                                      \end{equation}
Therefore, $S_{1}<1/\Gamma (\nu)$ when $t \to \infty$. In order to get the
lower approximation for the sum (for $1/2<\nu<1$) we shall multiply it by
$t$ and differentiate twice. Reducing the argument in the gamma-function
in the numerator by $\nu<1$ and increasing the denominator by $(2-2\nu)<1$
we then get
 \begin{eqnarray}
\nonumber \frac{d^{2}}{dt^{2}}(S_{1}t)&>&\sum_{k,l=1}^{\infty}
 \frac{(-\gamma)^{k+l}}{\Gamma (1+\nu k) \Gamma (\nu l)}
 \frac{\Gamma (\nu k+\nu l -1)}{ \Gamma (\nu k+\nu l+2)}  t^{\nu k+\nu l} \\
&&=\frac{1}{2}\frac{d}{dt}\Bigl[ \sum_{k=1}^{\infty}
 \frac{(-\gamma)^{k}t^{\nu k}}{\Gamma (1+\nu k)}
 \Bigr]^{2} \equiv \frac{1}{2}\frac{d}{dt}
 \Bigl[E_{1,\nu}(-\gamma t^{\nu})-1 \Bigr]^{2}
                                                           \end{eqnarray}
Since this inequality holds for every $t$, we can integrate the left-hand
and right-hand sides, which yields $S_{1}>1/2$ for $t \rightarrow \infty$.

\end{document}